\documentstyle[aps,12pt]{revtex}
\input psfig
\newcommand{\beq}[2]{\begin{equation}#1\label{#2}\end{equation}}
 \newcommand{\ceq}[1]{(\ref{#1})}
\newcommand{\mbd}[1]{\mbox{\bf #1}}
\newcommand{\bp}{\mbd{p}}
\newcommand{\bq}{\mbd{q}}

\newcommand{\bu}{\mbd{u}}
\newcommand{\br}{\mbd{r}}
\newcommand{\bj}{\mbd{j}}
\newcommand{\brs}{\mbox{\mbld r}}

\newcommand{\ba}{\mbd{a}}
\newcommand{\bb}{\mbd{b}}
\newcommand{\bc}{\mbd{c}}
\newcommand{\bd}{\mbd{d}}
\newcommand{\bA}{\mbd{A}}
\newcommand{\bD}{\mbd{D}}
\newcommand{\bB}{\mbd{B}}
\newcommand{\bC}{\mbd{C}}

\newcommand{\tgm}{G_{\{m\}}(\{\br\},\{L\};\{\br'\},0)}
\newcommand{\tgmm}{G_{\{m,M\}}(\{\br\},\{L\};\{\br'\},0)}

\newcommand{\tgl}{G_{\{\lambda\}}(\{\br\},\{L\};\{\br'\},0)}
\newcommand{\tgll}{G_{\{\lambda\Lambda\}}(\{\br\},\{L\};\{\br'\},0)}
\newcommand{\tgz}{G_{\{\lambda\}}(\{\br\},\{\br'\},\{ z\})}
\newcommand{\tgzll}{G_{\{\lambda,\Lambda\}}(\{\br\},\{\br'\},\{ z\})}
\newcommand{\ttgz}{G_\lambda(\{\br\},\{\br'\},\{ z\})}
\newcommand{\tgmz}{G_m(\{\br\},\{\br'\},\{ z\})}

\newcommand{\tglfc}{G(\br_i,L_i;\br_i',0|\phi_i,\bC^{(i)})}
\newcommand{\tgzfc}{G(\br_i,\br_i';z_i|\phi_i,\bC^{(i)})}

\newcommand{\Aij}{\bA^{(i)}_{(j)}}
\newcommand{\Bij}{\bB^{(i)}_{(j)}}
\newcommand{\aijk}[1]{a^{(i)}_{#1(jk)}}
\newcommand{\bjik}[1]{b^{(j)}_{#1(ik)}}
\newcommand{\ckij}[1]{c^{(k)}_{#1(ij)}}
\newcommand{\atijk}[1]{\tilde a^{(i)}_{#1(jk)}}
\newcommand{\btjik}[1]{\tilde b^{(j)}_{#1(ik)}}
\newcommand{\ctkij}[1]{\tilde c^{(k)}_{#1(ij)}}
\newfont{\mbld}{cmbx10 scaled 800}
\begin{document}
\title{FIELD THEORIES OF TOPOLOGICAL RANDOM WALKS}
\author{Franco Ferrari and Ignazio Lazzizzera\\
{\it Dipartimento di Fisica, Universit\'a di Trento, 38050 Povo (TN),
Italy.}\\
{and}\\
{\it INFN, Gruppo Collegato di Trento, Italy.}}
\date{June 99}
\maketitle
\vspace{-4.5in} \hfill{Preprint UTF XXX/YY} \vspace{4.4in}
\begin{abstract}
In this work we derive certain topological theories of transverse vector
fields whose amplitudes reproduce
topological invariants involving the interactions among the trajectories
of three and four random walks.
This result is applied to the construction of a field theoretical model
which describes the statistical
mechanics of an arbitrary number of topologically linked polymers
in the context of the analytical approach of Edwards.
With respect to previous attempts, our approach is very general, as
it can treat a system involving an arbitrary number of polymers
and the topological states are not only specified by the Gauss linking
number, but also by higher order topological invariants.
\end{abstract}
\section{Introduction}
Topologically linked
random chains  are studied in connection with
physical systems in which
the topological constraints of one-dimensional objects play an essential role
\cite{kleinertI},\cite{kleinertII}.
Examples are vortex rings in fluids and dislocation lines in solids.
To be concrete, the theory of random chains is applied here to the
description of polymers in a good solvent \cite{flory}--\cite{doiedw},
where the random chains are subjected to excluded volume interactions and
topological interactions \cite{kleinertI},\cite{necreview}--\cite{kholovil}.
The former take into account the effective
repulsions experienced by the polymers, while the latter arise
due to presence of stable topological constraints in the system.

In a recent publication \cite{ferlaz}, the so-called analytical approach
of Edwards \cite{edwardsI}
has been extended to the case of $N$ entangling polymers, mapping
the statistical mechanical problem of computing their free energy
to a field theoretical problem. The model obtained in this way is a
O$(n)$ field theory coupled to Chern-Simons (C-S) 
terms \cite{chernsimons} in the limit
$n\rightarrow 0$.
At one loop approximation, one finds that the topological interactions
tend to counterbalance the repulsive effects of the excluded volume forces.
This effect has been indeed observed in nature, for instance in the DNA
of some bacteria, which forms topologically entangled
rings \cite{ledobovi}.
Moreover, one can compute the second topological momentum
of two polymers exactly.
The analytical approach of Edwards is however limited by the fact that
the topological states of the system are distinguished using the
Gauss linking number. The latter is
a relatively poor topological invariant and 
describes topological interactions in which only the trajectories
of two polymers are involved.
Despite many efforts \cite{brsh}--\cite{moka}, the inclusion
in the treatment of topological random walks of more sophisticated
link invariants,
like for instance the Alexander and Jones polynomials \cite{kauffmann},
\cite{jones},
has not yet
been achieved.  The reason is that higher order link invariants
have no immediate
relation to the physical conformations of the polymers in the space
\cite{otvi}.
A remedy invoked by many authors is the introduction of C-S
field theories coupled to the polymer trajectories \cite{kleinertI},
\cite{kholovil}.
Unfortunately, this is not a simple task. First of all,
the amplitudes
of non-Abelian C-S field
theories contain topological invariants
\cite{witten}, but it is not clear how to use them to
impose constraints on the configurations of the system.
On the other side, it is not easy to find a
regularization or a mechanism that suitably removes the 
spurious non-topological contributions arising in C-S field theories.
For example, the introduction of a framing \cite{witten}
complicates the integrations
over the polymer trajectories to the extent that the mapping of the
statistical mechanics of the system to a field theory is no longer
possible \cite{felaone}.

To solve the above difficulties, we propose here interacting Abelian field
theories containing transverse vector fields.
Such theories are topological but not
gauge invariant. Of course, since the radiative corrections are not protected
by a gauge principle, counter-terms may arise which
are of a non-topological nature. To exclude this possibility, the couplings
among the vector fields are
chosen  so that
every quantum contribution disappears.
In this way one is able to construct theories whose amplitudes are purely
classical
and, in principle, can generate topological terms describing
interactions among an arbitrary number of
random chains. Apart from a constant factor, each
of these terms can be identified 
with the contribution of a tree-level
Feynman diagram appearing in a non-Abelian
C-S field theory \cite{agua}--\cite{gmmproc}.
However, it is exactly the freedom of choosing these factors
that allows the fixing of the topological costraints.
The above findings are used to build a model
of entangling polymers in which the topological interactions
induced by the Gauss linking invariant in the standard Edward
approach are corrected by higher order topological
interactions among three trajectories. The corresponding topological
invariant has a simple physical interpretation which is expressed 
in terms of magnetic fields generated by fictitious charged particles
moving along the trajectories of the polymers.
A generalization to the case of four trajectories is also outlined.

The material presented in this paper is divided as follows.
In the next Section we explain  the treatment of the $N$ polymers problem
of  \cite{ferlaz}
based on the use of the Gauss linking number to distinguish the
topological states. With respect to \cite{ferlaz}
some new investigations are made. For instance, the role of the
C-S fields as propagators of the collective modes which are relevant
in the topological entanglement \cite{vilbreav} is explored in details.
In Section three that approach is extended to include also higher
order topological interactions. To this purpose, abelian theories
are defined that generate topological interactions among three and
four loops.
Finally, the Conclusions are drawn in section four.
\section{The $N-$polymers problem}
Let $P_1,\ldots,P_N$ be a set of topologically entangling random chains
at thermal equilibrium. If the step length $a$ of the
segments composing the chains is very small, one can describe the
chains as trajectories in the space parametrized by vectors
$\br_i(s_i)$, $i=1,\ldots,N$ and continuous parameters
$s_1,\ldots,s_N$ such that:
\beq{0\le s_i\le
L_i\qquad\qquad\qquad 
\br_i(0)=\br'_i, \br_i(L_i)=\br_i} {trajectdef}
$L_i$ coincides with the contour length of $C_i$.
In first approximation the topological constraints will be imposed exploiting
the Gauss linking number:
\beq{
 \chi(C_i,C_j)\equiv 
 \frac{1}
 {4\pi}
 \int_0^{L_i}
 \int_0^{L_j}
 d\br_i(s_i)\times
 d\br_j(s_j)\cdot
    \frac
            {
              \left(
                     \br_i(s_i)-
                     \br_j(s_j)
              \right)
             }
             {
              |\br_i(s_i) - \br_j(s_j)|^3
             }
}
{glinv}
and requiring the conditions:
\beq{\chi(C_i,C_j) = m_{ij}}{topconstraints}
where $m_{ij}=m_{ji}$, $m_{ii}=0$ are a set of topological numbers.

To describe the statistical mechanics of the chains we define the
configuration probability $\tgm$.
This function measures the probability that the trajectories
$C_i$
have extrema (in the open case) at the points $\br_i'$ and $\br_i$
or a fixed point (in the closed case) in $\br_i'=\br_i$ for
$i=1,\ldots,N$.
Moreover, they should fulfill the topological conditions \ceq{topconstraints}.
In our notations $\{m\}$ denotes the $n\times n$ symmetric matrix of
topological numbers,
while $\{\br\}= \br_1,\ldots,\br_N$, $\{L\}=L_1,\ldots,L_N$ etc.
In the path integral approach
one obtains the
following expression of
$\tgm$ \cite{ferlaz}:
\beq{
\tgm =
\int_{\br_1'}^{\br_1}\ldots \int_{\br_N'}^{\br_N}
\mbox{\rm exp}\left\{-({\cal A}_0+{\cal A}_{ev})\right\}
\prod_{i=1}^{N-1}\prod_{j=2\atop j>i}^N\delta(\chi(C_i,C_j)-m_{ij})}
{gml}
where
\beq{
{\cal A}_0=\frac 3{2a}\sum_{i=1}^N\int_0^{L_i}\dot{\br}^2_i(s_i)}{azero}
is the action of a free random walk and
\beq{{\cal A}_{ev}=\frac 1{2a^2}\sum_{i,j=1}^N
\int_0^{L_i}
ds_i\int_0^{L_j}
ds'_jv^0_{ij}\delta^{(3)}(\br_i(s_i)-\br_j(s'_j))}{aev}
takes into account the excluded volume interactions.
For convenience, we have put:
\beq{
v^0_{ij}=
\left\{
\begin{array}{c c}
\tilde v^0_{ij}&\mbox{\rm for}\enskip i=j\cr
\tilde v^0_{ij}/2&\mbox{\rm for}\enskip i\ne j\cr
\end{array}
\right.\qquad\qquad \tilde v^0_{ij}=\tilde v^0_{ji}}{vzij}
where the $\tilde v^0_{ij}$ are coupling constants with the dimension of a
volume.
The Fourier transformed of $\tgm$ with respect to the parameters $m_{ij}$
is:
\beq{\tgl = \int_{\br_1'}^{\br_1}{\cal D} \br_1(s_1) \ldots
\int_{\br_N'}^{\br_N}{\cal D} \br_N(s_N)
\mbox{\rm exp}\left\{-\left( {\cal A}_0+{\cal A}_{ev}+{\cal A}_{2L}\right)
\right\}}{gll}
with
\beq{
{\cal A}_{2L}=i\sum_{i=1}^{N-1}\sum_{j=2\atop
j>i}^N\chi(C_i,C_j) \lambda_{ij}}{actop}
Following the approach of Edwards, one would like to transform the
above path integral into a field theoretical problem.
First of all, we treat the excluded volume interactions.
To this purpose, we introduce $N$ Gaussian scalar fields $\phi_1(\br),
\ldots,\phi_N(\br)$,
with action
\beq{{\cal A}_{\phi}=\frac{a^2}2\sum_{i,j=1}^N\int
d^3\br\phi_i[(v^0)^{-1}]^{ij}\phi_j}{aphi}
The fundamental identity which relates the excluded volume term
to a field theory amplitude is:
\beq{e^{-{\cal A}_{ev}}=\int{\cal D}\phi_1\ldots{\cal D}\phi_N\enskip
\mbox{\rm exp}\left\{-{\cal A}_{\{\phi\}} -i \sum_{i=1}^N\oint_{C_i}
ds_i \phi_i(\br_i(s_i))\right\}}{evfundid}
In the case of topological interactions one needs instead
Chern-Simons fields
$\Aij$ and $\Bij$ with action:
\beq{
S_{CS} =\frac \kappa{4\pi} \int d^3\br\sum_{i=1}^{N-1}
\sum_{j=2\atop j>i}^N \Aij\cdot(\nabla\times\Bij)}{csaction}
The C-S theory will be quantized in the Landau gauge, where the fields
are completely transverse.
In the coupling with the random chains only the following linear combinations
of fields are relevant:
\beq{
\bC^{(1)}=\sum\limits_{j=2}^N\frac k{4\pi}\bA^{(1)}_{(j)}}{cone}
\beq{
\bC^{(i)}=\sum\limits_{j=3\atop j>i}^N\frac k{4\pi}\bA^{(i)}_{(j)}
+\sum\limits_{j=1\atop j<i}^{N-2}\lambda_{ji}\bB^{(j)}_{(i)}
\qquad\qquad i=2,\ldots,N-1}{cigen}
and
\beq{
\bC^{(N)}=\sum\limits_{i=1}^{N-1}\lambda_{iN}\bB^{(i)}_{(N)}}{cenne}
The topological term appearing in the configurational
probability \ceq{gll}  can be rewritten  as an amplitude  of the above
C-S field theory as follows:
\beq{
\int{\cal D\bA}{\cal D\bB}\enskip\mbox{\rm exp}\left\{-i S_{CS}-i\sum_{i=1}^N
\int_0^{L_i}\bC^{(i)}(\br(s_i))d\br(s_i)\right\}=e^{-{\cal A}_{2L}}}
{topfundid}
where
\beq{{\int\cal D\bA}{\cal D\bB}\equiv\int\prod_{i<j=1}^N
{\cal D}\Aij{\cal D}\Bij}
{csmeas}
\begin{figure}
\vspace{1.5truein}
\includegraphics{twoloopint.eps}
\vspace{.2in}
\caption{Graphical interpretations of the topological interactions
between two random walks $C_i$ and $C_j$
mediated by the Chern-Simons fields $\Aij,\Bij$.}
\label{twoloopintgraph}
\end{figure}
Exploiting the identities \ceq{evfundid} and \ceq{topfundid}
the configurational probability \ceq{gll} becomes:
\beq{
\tgl=\langle\prod_{i=1}^N\tglfc\rangle_{\{\phi\},\{\bA\},\{\bB\}}}
{glfact}
where
\beq{
\tglfc=\int_{\br'_i}^{\br_i}
{\cal D}\br_i(s_i)
e^{-\int_0^{L_i}{\cal L}_ids_i}}{glfrr}
and
\beq{{\cal L}_i=
\frac 3{2a}
\dot{\br}_i^2(s_i)+i\phi_i(\br_i)
-i\dot{\br}_i(s_i)
\cdot\bC^{(i)}(\br_i(s_i))
}
{lagem}
As we see from eqs. \ceq{glfact} the trajectories
$C_1,\ldots,C_N$  are completely decoupled before averaging
over the auxiliary fields $\{\phi\},\{\bA\},\{\bB\}$.
Formally, 
$\tglfc$ is the evolution kernel of the
random walk of a particle in an
electromagnetic field $(i\phi_i,\bC^{(i)})$.
Thus, it satisfies the pseudo-Schr\"odinger equation \cite{tanaka}:
\begin{equation}
\left[\frac\partial{\partial{L_i}}
-\frac a 6\mbd{D}_i^2+i\phi_i\right]
G(\br_i, L_i;\br'_i,0|\phi_i,\bC^{(i)})=
\delta(L_i)\delta(\br_i-\br'_i)
\label{pschroedeq}
\end{equation}
The covariant derivatives $\bD_i$ appearing in the above equation
are given by:
\beq{\bD_i=\nabla+i\bC^{(i)}\qquad\qquad\qquad i=1,\ldots,N}{covdevdef}
It is now convenient to  perform a Laplace transformation of
$\tglfc$ with respect to the length $L_i$:
\beq{
\tgzfc=\int_0^\infty dL_ie^{-z_iL_i}\tglfc}{evka}
Accordingly, we are now considering the Laplace transformed
configurational probability:
\beq{
\tgz=\int_0^{+\infty}dL_1\ldots\int_0^{+\infty}dL_N
\mbox{\rm exp}\left\{-\sum_{i=1}^Nz_iL_i\right\}\tgl}{laplabig}
Since the order of the integrations over the auxiliary fields and
the lengths $L_i$ of the trajectories can be permuted, we have:
\beq{\tgz=\langle\prod_{i=1}^N\tgzfc\rangle_{\{\phi\},\{\bA\},\{\bB\}}}
{facttwo}
The new variables $z_i$ play the role of  Boltzmann-like factors which
govern the distribution lengths of the random chains.
The advantage of having performed the Laplace transformations is that
$\tgzfc$ satisfies a $stationary$ pseudo Schr\"odinger equation:
\beq{\left[z_i-H_i
\right]
\tgzfc=\delta(\br_i-\br'_i)}{stapseschroed}
Here the Hamiltonian $H_i$ is given by:
\beq{H_i=\frac a 6\bD_i^2-i\phi_i}{glhamil}
The solution of eq. \ceq{stapseschroed} can now be expressed in terms of
second quantized fields fields $\psi_i^*,\psi_i$, $i=1,\ldots,N$:
\begin{equation}
\tgzfc=
{\cal Z}_i^{-1}\int
{\cal D}\psi_i
{\cal D}\psi^*_i
\psi_i(\br_i)
\psi^*_i(\br'_i)
e^{-F[\psi_i]}
\label{glprop}
\end{equation}
where
$F[\psi_i]$ represents the Ginzburg-Landau free
energy of a superconductor in a fluctuating magnetic field:
\begin{equation}
F[\psi_i]=\int d^3\br\left[
\frac a 6|{\mbd D}_i\psi_i|^2 +
(z_i+i\phi_i)|\psi_i|^2
\right]
\label{glfen}
\end{equation}
and ${\cal Z}_i$ is the partition function of the system:
\begin{equation}
{\cal Z}_i=\int
{\cal D}\psi_i
{\cal D}\psi^*_i
e^{-F[\psi_i]}
\label{ztpartfun}
\end{equation}
The auxiliary fields $\phi_i$
can be eliminated from the configurational probability \ceq{facttwo}
integrating them out. The integration over these fields is non-trivial  
due to the presence of the factors ${\cal Z}_i^{-1}$ in the
second quantized expression of
$\tgzfc$ in \ceq{glprop}, but can be made Gaussian by
exploiting the identity \cite{repltrick}:
\beq{{\cal Z}_i^{-1}=\lim_{n_i\to 0}{\cal Z}_i^{n_i-1}}{replide}
In this way
\beq{\tgzfc==\lim_{n_i\to 0}{\cal Z}_i^{n_i-1}
\int
{\cal D}\psi_i
{\cal D}\psi^*_i
\psi_i(\br_i)
\psi^*_i(\br'_i)
e^{-F[\psi_i]}}{replmeth}
The above equation should be understood as follows:
the right hand side is first computed supposing that the replica index $n_i$ is
an arbitrary positive integer and then one performs the analytic continuation
of the result to the point $n_i=0$.
Now $\tgzfc$ is a product of $n_i$ path integrals. To each one we associate
a set of replica fields $\psi_i^{a_i}$, $a_i=1,\ldots,n_i$,
so that it will be convenient to introduce the multiplets:
\beq{\Psi_i=(\psi_i^1,\ldots,\psi_i^{n_i})}{psimult}
\beq{\Psi_i^*=(\psi_i^{*1},\ldots,\psi_i^{*n_i})}{psibmult}
and to rewrite $\tgzfc$ as follows:
\begin{equation}
\tgzfc=
\lim_{n_i\to 0}\int
{\cal D} \Psi_i
{\cal D} \Psi_i^*
\psi_i^1(\br_{i})
\psi_i^{*1}
(\br'_i)\enskip
e^{-F[\Psi_i]}
\label{pgl}
\end{equation}
where
\[ F[\Psi_i]=\sum_{a_i=2}^{n_i}\int d^3\br\left[
\frac a 6|\bD\psi_i^{a_i}|^2+(z_i+i\phi_i)|\psi_i^{a_i}|^2\right]
\]
\begin{equation}
\equiv
\int d^3\br\left[
\frac a 6|\bD\Psi_i|^2+(z_i+i\phi_i)|\Psi_i|^2\right]
\label{freearepl}
\end{equation}
and
\beq{\int
{\cal D} \Psi_i
{\cal D} \Psi_i^*\equiv
\prod_{a_i=1}^{n_i}{\cal D}\psi_i^{a_i}{\cal D}\psi_i^{*a_i}}{psipsimeas}
According to the replica method, one supposes
that it is possible to commute the limit of vanishing replica index
with the path integrations over the auxiliary fields.
Thus,
substituting eq. \ceq{pgl} in \ceq{facttwo} and
performing
the integration over the scalars
$\phi_i$, which is now Gaussian, one obtains the
final expression of the configurational
probability $\tgz$:
\[
\tgz=
\]
\begin{equation}
\lim_{n_1,\ldots,{n_N\to 0}}\int
{\cal D} \Psi
{\cal D} \Psi^*
\int{\cal D}\bA{\cal D}\bB
\prod_{j=1}^N\left[\psi_j^1(\br_j)
\psi_j^{*1}(\br'_j)
\right]\enskip\mbox{\rm exp}
\left\{-{\cal A}_{Gauss}
\right\}
\label{prefin}
\end{equation}
${\cal A}_{Gauss}$ is the free energy of the topologically linked random
walk written in terms of C-S and second quantized fields. The subscript
refers to the fact that the topological constraints have been imposed using the
Gauss linking number. After a rescaling the complex scalar fields
of the kind
\beq{\Psi_i\rightarrow\sqrt{\frac M2}\Psi_i\qquad\qquad\qquad\Psi_i^*
\rightarrow\sqrt{\frac M2}\Psi_i^*}{rescaling}
where $M$ is a mass parameter, the explicit expression of ${\cal A}_{Gauss}$
is:
\[{\cal A}_{Gauss}=iS_{CS}
+\sum_{i=1}^N\int d^3\br \left[|
\bD_i\Psi_i|^2+m_i^2|\Psi_i|^2\right]+
\]
\beq{
\sum_{i,j=1}^N \frac {2M^2v^0_{ij}}{a^2}
\int d^3\br |\Psi_i|^2|\Psi_j|^2}{finalfen}
where
\beq{m_i^2=2Mz_i}{massdefs}
The above  action describes  a O$(n)$  model coupled  to  Chern-Simons
fields in the limit $n=0$. 
The analogous of the Planck constant is here the constant
$\hbar=\frac {Ma}3$ and has been set equal to one in \ceq{finalfen}.
The topological fields are not just auxiliary, but play a physical role,
since they propagate the long-range interactions that
impose the topological constraints \ceq{topconstraints}. 
One may argue at this point that the number of C-S fields used in the present
approach, which is $N(N-1)$, is highly redundant with respect to the physical
number of degrees of freedom involved if $N>3$. This number can be
computed exploiting
refs.  \cite{vilbreav},\cite{brereton},  where  a   set of  collective
modes which   are relevant in  the  topological interactions has been
constructed in terms of the so-called bond vector
densities:
\beq{\bu_i(\br)=\oint_{C_i}d\br_i\delta(\br-\br_i)\qquad\qquad\qquad
i=1,\ldots,N}{bondvecdens}
More precisely, the collective  modes  are linear combinations  of the
$N$ bond vector
densities in the Fourier space, where
\beq{\bu_i(\bq)=\oint_{C_i}d\br_ie^{1\bq\cdot\br_i}\qquad\qquad\qquad
i=1,\ldots,N}{frbonvecden}
Indeed, the Gauss linking invariant
\ceq{glinv} can be expressed as follows:
\beq{\chi(C_i,C_j)=\int \frac{d^3\bq}{\bq^2}\enskip\bq\cdot(\bu^i(\bq)\times
\bu^j(-\bq))}{ginvrelmod}
As a consequence, if there is a number $M\le N$ of random
chains which have non-trivial topological relations with the others,
i.    e. the   maximum  rank   of   the    matrices $\{m_{ij}\}$   and
$\{\lambda_{ij}\}$ is $M$, there are at most $M$ degrees of freedom
to be propagated.
Of course, this
is not  in contradiction with  our result.  As a matter  of
fact it is possible  to see that,  exploiting the equations  of
motions, the number of C-S fields in  the action \ceq{finalfen} can be
reduced to $N$.  However, the C-S theory obtained in this way is not
universal,  since it cannot describe  all the  topological states of the
system. The reason is that after the reduction the C-S propagators
depend  on the parameters  $\lambda_{ij}$ and
become singular in the limit in which some of them vanish. In general,
it has not been possible to build a  suitable Abelian C-S field theory
with  less than $N(N-1)$ fields  without  encountering the problem of
diverging propagators wherever $rank[\lambda]<N$ or without resorting
to a complicated
parameterization of the matrix $\{\lambda\}$ provided for instance
by the solution of the following
algebraic system of equations:
\beq{\lambda_{ij}=\sum_{k=1}^N\eta_{ik}\eta_{kj}}{quadrsys} 

\section{Including Higher Loop Interactions}
The Gauss linking number describes a topological interaction between two
loops and it is quite a poor topological invariants.
Thus it would be interesting to include in the above approach also higher
order topological interactions.
To begin, we consider an interaction $\Gamma_3(C_i,C_j,C_k)$ among three
loops, where $i<j<k$. To determine
$\Gamma_3(C_i,C_j,C_k)$ we construct a suitable topological field theory
with action:
\[S_3(i,j,k)=\epsilon^{\mu\nu\rho}
\int d^3x\aijk{\mu}\partial_\nu\atijk{\rho}+
\epsilon^{\mu\nu\rho}
\int d^3x\bjik{\mu}\partial_\nu\btjik{\rho}\]
\beq{\epsilon_{\mu\nu\rho}
\int d^3x\ckij{\mu}\partial_\nu\ctkij{\rho}+
\Lambda^k_{ij}\epsilon_{\mu\nu\rho}\int d^3x\aijk{\mu}\bjik{\nu}\ckij{\rho}
}{csthree}
where the fields $\aijk{\mu},\ldots,\ctkij{\mu}$ are purely transverse.
$S_3(i,j,k)$ describes at the classical level
a topological field theory which is not gauge invariant.
Moreover, it is easy to convince oneself that the theory has no
radiative corrections that could spoil its topological
properties.
Despite of this fact, there are nontrivial amplitudes as for instance
the following correlator:
\beq{G_{\Lambda^i_{jk}}(C_i,C_j,C_k)=\langle
e^{ik\oint_{C_i}
d\brs_i^\alpha \atijk{\alpha}(\brs_i)}
e^{ik\oint_{C_j}
d\brs_j^\beta \btjik{\beta}(\brs_j)}
e^{i\gamma_3\oint_{C_k}
d\brs_k^\gamma \ctkij{\gamma}(\brs_k)}\rangle}{ftgt}
The above amplitude can be exactly computed and one obtains:
\beq{G_{\Lambda^i_{jk}}(C_i,C_j,C_k)=\mbox{\rm exp}\left\{
\Lambda^i_{jk}\Gamma(C_i,C_j,C_k)\right\}}{tgenres}
where
\beq{\Gamma(C_i,C_j,C_k)=\oint_{C_i}d\br_i^\alpha
\oint_{C_j}d\br_j^\beta\oint_{C_k}d\br_k^\gamma
I_{\alpha\beta\gamma}(\br_i,\br_j,\br_k)}{ttinv}
and
\beq{I_{\alpha\beta\gamma}(\br_i,\br_j,\br_k)=\epsilon^{\mu\nu\rho}
\int d^3\br G_{\mu\alpha}(\br-\br_i)G_{\nu\beta}(\br-\br_j)
G_{\rho\gamma}(\br-\br_k)}
{integrand}
In the above equation
\beq{ G_{\mu\nu}(\br_1-\br_2)=-\epsilon_{\mu\nu\rho}
\frac{(r_1-r_2)^\rho}{|\br_1-\br_2|^3}}{cprop}
$r^\mu$ being the components of the vector $\br$.
An analogous of $\Gamma(C_i,C_j,C_k)$ in the case $C_i=C_j=C_k$
has been studied in connection with perturbative calculations of
self-linking invariants in non-Abelian C-S field theories.
After a volume integration in \ceq{integrand}, the right hand side
of eq. \ceq{ttinv} has a complicated expression, which has been evaluated
in \cite{gmmproc}.
With respect to \cite{gmmproc}, however, one does not need
path ordering of the trajectories, so that it is possible to perform
the volume integration
using a different strategy.
To this purpose, let us define the currents:
\beq{j^\mu_{(l)}(\br)\equiv\oint_{C_l}d\br^\mu_l\delta(\br-\br_l)}{currdef}
and the magnetic fields:
\beq{{\cal B}_{(l)\mu}(\br)=\epsilon_{\mu\nu\rho}\oint_{C_l}d\br^\nu
\frac{(r-r_l)^\rho}
{|\br-\br_l|^3}}{magfiedef}
with $\nabla\cdot\vec{\cal B}_{(l)}=0$ and
$\nabla\times\vec{\cal B}_{(l)}=4\pi \bj_{(l)}$.
The vector potentials corresponding to these magnetic fields are:
\beq{\vec{\cal A}_{(l)}(\br)=\oint_{C_l}\frac {d\br_l}{|\br-\br_l|^3}=
\int d^3\br'\frac{\bj_{(l)}(\br')}{|\br-\br'|}}{vecpotfdef}
ad satisfy the relations
\beq{
\nabla\cdot\vec{\cal A}_{(l)}=0\qquad\qquad\qquad
\nabla\times\vec{\cal A}_{(l)}=
\vec{\cal B}_{(l)}}{potrelone}
\beq{\triangle\vec{\cal A}_{(l)}=-4\pi \bj_{(l)}
}{potreltwo}
In terms of the magnetic fields ${\cal B}^\mu_{(l)}$ we have:
\beq{\Gamma(C_i,C_j,C_k)=\int d^3\br
\vec{\cal B}_{(i)}(\br)\cdot\vec{\cal B}_{(j)}(\br)
\times\vec{\cal B}^\rho_{(k)}(\br)
}{gcsecexp}
The space integral can be eliminated using the Stokes theorem and after
some calculations one obtains:
\[\frac 1{4\pi}\Gamma(C_i,C_j,C_k)=-\int\int_{\Sigma_{C_k}}d{\bf S}_k\cdot
\vec{\cal B}_{(i)}(\br_k)\times\vec{\cal B}_{(j)}(\br_k)+
\int\int_{\Sigma_{C_j}}d{\bf S}_j\cdot
\vec{\cal B}_{(i)}(\br_j)\times\vec{\cal B}_{(k)}(\br_j)\]
\beq{-\int\int_{\Sigma_{C_i}}d{\bf S}_i\cdot
\vec{\cal B}_{(j)}(\br_i)\times\vec{\cal B}_{(k)}(\br_i)}{gcthiexp}
where the $\Sigma_{C_i}, \Sigma_{C_j},\Sigma_{C_k}$ are arbitrary surfaces
having respectively $C_i,C_j,C_k$ as
borders and infinitesimal surface elements $d{\bf S}_i,d{\bf S}_j,d{\bf S}_k$.
To include the interactions among the loop trajectories,
we extend the previous configurational probability of eq. \ceq{gml} as
follows:
\[
\tgmm =
\int_{\br_1'}^{\br_1}\ldots \int_{\br_N'}^{\br_N}
\mbox{\rm exp}\left\{-({\cal A}_0+{\cal A}_{ev})\right\}\]
\beq{
\prod_{i=1}^{N-1}\prod_{j=2\atop j>i}^N\delta(\chi(C_i,C_j)-m_{ij})
\prod_{i=1}^{N-2}\prod_{j=2\atop j>i}^{N-1}
\prod_{k=3\atop k>j}^N
\delta(\Gamma(C_i,C_j,C_k)-M_{ijk})}
{gmml}
After performing the Fourier transformations with respect to the
topological numbers $m_{ij}$ and $M_{ijk}$ one obtains:
\beq{
\tgll = \int_{\br_1'}^{\br_1}{\cal D} \br_1(s_1) \ldots
\int_{\br_N'}^{\br_N}{\cal D} \br_N(s_N)
\mbox{\rm exp}\left\{-\left( {\cal A}_0+{\cal A}_{ev}+{\cal A}_{2L}
+{\cal A}_{3L}\right)
\right\}}{glll}
with
\beq{
{\cal A}_{3L}=
i\sum_{i,j,k=1\atop i<j<k}^N\left\{
\Lambda^i_{jk}\Gamma(C_i,C_j,C_k)\right\}}{tlaction}
At this point
we rewrite the higher
loop topological interactions appearing in \ceq{glll} as a field theory
amplitude using eqs. \ceq{tgenres} and \ceq{ttinv}:
\[\int {\cal D}[a]{\cal D}[b]{\cal D}[c]
{\cal D}[\tilde a]{\cal D}[\tilde b]{\cal D}[\tilde c]e^{-iS_{3L}}\]
\beq{
\left[
\prod_{i,j,k=1\atop i<j<k}^N
e^{ik\oint_{C_i}
d\brs_i^\alpha \atijk{\alpha}(\brs_i)}
e^{ik\oint_{C_j}
d\brs_j^\beta \btjik{\beta}(\brs_j)}
e^{i\gamma_3\oint_{C_k}
d\brs_k^\gamma \ctkij{\gamma}(\brs_k)}\right]=
e^{-i{\cal A}_{3L}}
}
{tlfund}
${\cal D}[a],\ldots,{\cal D}[\tilde c]$ denotes the measure
over the  C-S fields
$\aijk{\mu},\ldots,\ctkij{\mu}$ and
\beq{S_{3L}=\sum_{i,j,k=1\atop i<j<k}^NS_3(i,j,k)}
{stl}
Again, the C-S fields decouple the topological interactions.
The integration over the trajectories of the random chains can
now be performed following the same strategy of the
previous section. In this way one obtains the following expression of the
configurational probability in terms of the Laplace variables
$z_1,\ldots,z_N$:
\[\tgzll=\]
\beq{
\lim_{n_1,\ldots,{n_N\to 0}}
{\cal D} \Psi
{\cal D} \Psi^*
\int{\cal D}\bA{\cal D}\bB
{\cal D}[a]\ldots{\cal D}[\tilde c]
\prod_{i=1}^N\left[\psi_i^1(\br_i)\psi^{*1}_i(\br_i)'\right]
e^{-{\cal A}_h}}
{tlfinpi}
where
\[{\cal A}_h=iS_{CS}+iS_{3L}
+\sum_{s=1}^N\int d^3\br \left[|
\bD_s^{ext}\Psi_s|^2+m_s^2|\Psi_s|^2\right]+
\]
\beq{
\sum_{i,j=1}^N \frac {2M^2v^0_{ij}}{a^2}
\int d^3\br |\Psi_i|^2|\Psi_j|^2}{tlfinalfen}
The covariant derivatives that include the three loop interactions
are given by:
\beq{
\bD_s^{ext}=
\nabla
+i
(
  \bC^{(s)}
  +\tilde{\bd}^{(s)})}{extcovdev}
with
\beq{\tilde{\bd}^{(s)}=\sum_{j,k=1\atop s<j<k}^N\tilde{\ba}^{(s)}_{(jk)}+
\sum_{i,k=1\atop i<s<k}^N\tilde{\bb}^{(s)}_{(ik)}+
\sum_{i,j=1\atop i<j<s}^N\tilde{\bc}^{(s)}_{(ij)}}{dts}
We notice that within  the above approach  it  is possible  to include
also topological interactions among four or
more trajectories. Let us consider for instance a
topological four loop interaction $\Gamma(C_1,C_2,C_3,C_4)$. 
To generate such interaction we define the following action:
\[
S_4(1,2,3,4)=
{\kappa\over 4\pi}\sum_{i=1}^5\int d^3x\epsilon^{\mu\nu\rho}
a_\mu^{(i)}\partial_\nu b^{(i)}_\rho+\]
\beq{
\Lambda_1\int d^3x\epsilon^{\mu\nu\rho}
a^{(1)}_\mu a^{(5)}_\nu a^{(2)}_\rho+
\Lambda_2\int d^3x\epsilon^{\mu\nu\rho}
a^{(3)}_\mu b^{(5)}_\nu a^{(4)}_\rho}{fliact}
As in the previous case, the theory is topological but has no gauge
invariance. 
This  can be  dangerous, since radiative  corrections  may arise which
spoil the topological properties of the theory, but it  is easy to see
that the above theory has no quantum contributions.

The relevant correlation function to be considered here is:
\beq{G_{\Lambda_1\Lambda_2}(\gamma_1,\gamma_2,\gamma_3,\gamma_4)=
\langle\prod_{i=1}^4 e^{i\gamma_i\oint_{C_i}dx_i^{\mu_i}b^{(i)}_{\mu_i}}
\rangle}{giiglogt}
The above amplitude can be exactly computed
and the result is:
\beq{G_{\Lambda_1\Lambda_2}(\gamma_1,\gamma_2,\gamma_3,\gamma_4)=
\mbox{\rm exp}\left\{-i\Lambda_1\Lambda_2\left[\prod_{i=1}^4
\gamma_i\right]\Gamma(C_1,C_2,C_3,C_4)\right\}}{gllgf}
where
\[
\Gamma(C_1,C_2,C_3,C_4)=\prod_{i=1}^4\left[\oint_{C_i}dx_i^{\alpha_i}\right]
\int d^3x \int d^3y
\epsilon^{\lambda\mu\nu}\epsilon^{\rho\sigma\tau}
\]
\beq{
\times G_{\mu\sigma}(x-y)G_{\lambda\alpha_1}(x-x_1)G_{\nu\alpha_2}(x-x_2)
G_{\rho\alpha_3}(y-x_3)G_{\tau\alpha_4}(y-x_4)}{gfour}
Eq. \ceq{gfour} describes a topological interaction among four loops.
Unfortunately, the elimination of the double volume integral
has not been possible following the strategy of Section II.
For this reason, an expression of $\Gamma(C_1,C_2,C_3,C_4)$ in terms of the
magnetic fields \ceq{magfiedef} could not be derived.
\section{Conclusions}
In the first part of this work the statistical mechanical problem of a
system of $N$ polymers whose topological interactions are governed by the
Gauss linking number has been mapped to a field theory following ref.
\cite{ferlaz}.
In the model obtained in this way the C-S fields play a physical role, since
they mediate the topological forces which impose the constraints
\ceq{topconstraints}.
We notice that the relation \ceq{topfundid}, which expresses the
topological contributions appearing in the {\it first quantized}
version of the configurational probability \ceq{gll} in terms of field
amplitudes, can be
reproduced also by means of other Abelian C-S field theories.
Each of these theories differs
from the other by the number of fields. However, the requirement that their
propagators should not diverge when $rank[\lambda]<N$ implies that the
independent  fields should be at least $N(N-1)$.
On the other side, if one starts
with $\bar N> N(N-1)$ C-S fields, it is always possible to reduce
their number to $N(N-1)$
by exploiting the equations of motion.
For this reason, the model of topologically entangling polymers given by eqs.
\ceq{prefin}--\ceq{massdefs} is unique.

In Section II the field theories are  essential in order to decouple
the random chains and to rewrite the
configurational probability in a {\it second quantized} form.
However, they start to
play an even more active role in Section III. In fact, here the
field  amplitudes are fundamental to provide the
explicit expression of the
topological interactions
among three or more polymer trajectories.
In the case of a three loop interaction, the topological invariant
$\Gamma(C_i,C_j,C_k)$ has also a nice physical interpretation
in terms of magnetic fields given by eq. \ceq{gcthiexp} as the
Gauss linking invariant.
With respect to non-Abelian C-S field theories,
the advantage of the theories of transverse vector fields
defined here is the possibility of generating
single  topological invariants $\Gamma(C_i,C_j,C_k)$ without the problem
of spurious non-topological contributions or of higher order corrections.
Moreover, the
freedom to choose the parameters $\Lambda^i_{jk}$
is crucial in order to
impose constraints on the $\Gamma(C_i,C_j,C_k)$ as shown by eqs.
\ceq{gmml} and \ceq{glll}.

Concluding, the field theoretical approach illustrated in this paper solves
in principle the most serious drawback present in the
analytical approach of Edwards, i. e. the
use of the Gauss linking invariant to specify the topological
status of a system of random chains.
However, more work and, possibly, numerical simulations are still needed
in order to evaluate the phenomenological implications of the new
topological terms introduced in the configurational probability
\ceq{tlfinpi} and to make contact with experiments.


\begin{thebibliography}{99}
\bibitem{kleinertI} H. Kleinert, {\it Path Integrals} (2nd edition),
World Scientific, Singapore, New Jersey, London, Hong Kong 1995.
\bibitem{kleinertII} H. Kleinert, {\it Gauge Fields in Condensed Matter},
Vol. 1, World Scientific, Singapore 1989.
\bibitem{flory}, {\it Principles of Polymer Chemistry}, Ithaca, N. Y. 1967.
\bibitem{degennesI}, {\it Scaling Concepts in Polymer Physics},
Cornell University Press, Ithaca, N.Y., 1979.
\bibitem{doiedw}
M. Doi and S. F. Edwards, {\it The Theory of Polymer Dynamics},
Clarendon Press, Oxford, 1986.
\bibitem{necreview}
S. Nechaev, {\it Int. Jour. Mod. Phys.} {\bf B4} (1990), 1809.
\bibitem{kholovil} A. L. Kholodenko and T. A. Vilgis,
{\it Phys. Rep.} {\bf 298} (1998), 251.
\bibitem{ferlaz} F. Ferrari and I. Lazzizzera, {\it Towards a Field Theoretical
Description of Topologically Linked Polymers}, hep-th/9903084.
\bibitem{edwardsI} S. F. Edwards, {\it Proc. Phys. Soc.} {\bf 91} (1967), 513;
 {\it J. Phys. A: Math. Gen.} {\bf 1} (1968), 15.
\bibitem{chernsimons}
R. Jackiw and S. Templeton, {\it Phys. Rev.} {\bf D23} (1981),
2291; S. Deser, R. Jackiw and S. Templeton, {\it Phys. Rev. Lett.}
{\bf 48} (1983), 975;
J. Schonfeld, {\it Nucl. Phys.} {\bf B185} (1981), 157;
C. R. Hagen, {\it Ann. Phys.} (NY) {\bf 157} (1984), 342.
\bibitem{ledobovi} S. D. Levene, C. Donahue, T. C. Boles and
N. R. Cozzarelli, {\it Biophys. Jour.} {\bf 69} (1995), 1036;
M. Otto and T. A. Vilgis, {\it Phys. Rev. Lett.} {\bf 80} (4) (1998), 881.
\bibitem{brsh} M. G. Brereton and S. Shah, {\it J. Phys. A: Math. Gen.} 
{\bf 13} (1980), 2751.
\bibitem{bresha} M. G. Brereton and S. Shah, {\it J. Phys. A: Math. Gen.}
{\bf 14} (1981), L-51; {\it ibid.}
{\bf 15} (1982), 989.
\bibitem{tanaka} F. Tanaka, {\it Prog. Theor. Phys.} {\bf 68} (1982), 148.
\bibitem{tanakaii}
F. Tanaka,
{\it Prog. Theor. Phys.} {\bf 68} (1982), 164.
\bibitem{otherworks}
D. J. Elderfield,  {\it J. Phys. A: Math. Gen.} {\bf 15} (1982), 1369;
J. Cardy, {\it Phys. Rev. Lett.} {\bf 72} (1994), 1580;
\bibitem{otvi} M. Otto and T. A. Vilgis,
{\it J. Phys. A: Math. Gen.} {\bf 29} (1996), 3893.
\bibitem{nero}
S. K. Nechaev and V. G. Rostiashvili, {\it J. Phys. II} {\bf 3} (1993), 91;
V. G. Rostiashvili, S. K. Nechaev and T. A. Vilgis, {\it Phys. Rev.}
{\bf E48} (5) (1993), 3314.
\bibitem{vilbreav} M. G. Brereton and
T. A. Vilgis, {\it Jour. Phys. A: Math. Gen.} {\bf 28} (1995), 1149.
\bibitem{brereton} M. G. Brereton, {\it Jour. Mol. Struct.} (Theochem),
{\bf 336} (1995), 191.
\bibitem{moka}
J. D. Moroz and
R. D. Kamien, {\it Nucl. Phys.} {\bf B506 [FS]} (1997), 695.
\bibitem{kauffmann} L. H. Kauffmann, {\it Knots and Physics}, World Scientific,
Singapore 1993.
\bibitem{jones} V. F. R. Jones, {\it Bull. Am. Math. Soc.} {\bf 129} (1989),
103.
\bibitem{witten} E. Witten, {\it Comm. Math. Phys.} {\bf 121} (1989), 351. 
\bibitem{felaone} F. Ferrari and I. Lazzizzera,
{\it Jour. Phys.} {\bf A}: {\it Math. Gen.} {\bf 32} (1999), 1347, 
hep-th/9803008.
\bibitem{agua}
E. Guadagnini, M. Martellini and M. Mintchev, {\it Nucl. Phys.}
{\bf B336} (1990), 581;
J. M. F. Labastida and A. V. Ramallo, {\it Phys. Lett.}
{\bf 238B} (1989), 214.
\bibitem{gmmproc} E. Guadagnini, M. Martellini and M. Mintchev,
{\it Chern-Simons Filed Theory and Link Invariants},
Talk presented at the 13$^th$ Johns Hopkins Workshop on
{\it Knots, Topology and Field Theory}, Florence, June 1989,
Preprint CERN-TH.5479/98.
\bibitem{repltrick} S. F. Edwards and P. W. Anderson,
{\it J. Phys.} {\bf F5} (1975), 965.
\bibitem{cowe} S. Coleman and E. Weinberg, {\it Phys. Rev.} {\bf D7}
(6) (1973), 1888.
\end{thebibliography}
\end{document}